\documentclass[runningheads]{llncs}

\usepackage[comma,authoryear,round,longnamesfirst,sectionbib]{natbib}

\usepackage{graphicx}
\usepackage{wrapfig}
\usepackage{booktabs}
\usepackage{amsmath,amsfonts,amssymb}
\usepackage{stmaryrd}

\usepackage{turnstile}
\usepackage{mathrsfs}
\usepackage{enumitem}
\usepackage{amsthm}
\usepackage[super]{nth}
\usepackage[toc,page]{appendix}
\usepackage{listings}
\usepackage{mathtools}
\usepackage{csquotes}
\usepackage{multicol}
\usepackage{fancyvrb}
\usepackage{makecell}
\usepackage{mdframed}
\usepackage[toc,page]{appendix}
\usepackage{mathdots}
\usepackage{empheq}
\usepackage{appendix}
\usepackage{bm}
	\newlength{\wordlength}

\usepackage{algorithm,algorithmic}

\usepackage{pgfplots}

\usepackage{array}
\newcolumntype{L}[1]{>{\raggedright\let\newline\\\arraybackslash\hspace{0pt}}m{#1}}
\newcolumntype{C}[1]{>{\centering\let\newline\\\arraybackslash\hspace{0pt}}m{#1}}
\newcolumntype{R}[1]{>{\raggedleft\let\newline\\\arraybackslash\hspace{0pt}}m{#1}}

\usepackage{tikz}
\usetikzlibrary{automata, positioning, calc, shapes, arrows, fit}

\usepackage[unicode=true, bookmarks=false, breaklinks=true, pdfborder={0 0 1}, colorlinks=true, citecolor=blue]{hyperref}

\usepackage{pifont}
\definecolor{MyGreen}{rgb}{0, 0.7, 0}
\definecolor{MyRed}{rgb}{0.8, 0, 0}

\newcommand{\V}{\mathcal{V}}

\newcommand{\A}{A}
\renewcommand{\G}{\mathcal{G}}
\newcommand{\E}{\mathcal{E}}

\newcommand{\Q}{\mathbb{Q}}

\newcommand{\tup}[1]{\langle #1 \rangle}

			\usepackage{boxedminipage}
			\usepackage{xspace}

\title{Individually Rational Land and Neighbor Allocation: Impossibility Results}

\author{Haris Aziz\inst{1,}\inst{2}}

\institute{UNSW Sydney, Australia  \\
\email{haris.aziz@unsw.edu.au}
\and
Data61 CSIRO
}

\makeatletter
\newcommand{\@chapapp}{\relax}%
\makeatother

\begin{document}
	\maketitle

	\begin{abstract}
		We consider a setting in which agents are allocated land plots and they have additive preferences over which plot they get and who their neighbor is. Strategyproofness, Pareto optimality, and individual rationality are three fundamental properties in economic design. We present two impossibility results showing that the three properties are incompatible in this context.	
	\end{abstract}

\section{Introduction}

We consider an allocation problem in which agents wish to get one of the plots and have preferences over them. They additionally have preferences over who their neighbor is. 
Such a problem was recently studied by \citet{EPTZ20a} who among other results, explored the issue of maximizing social welfare. 

The model of \citet{EPTZ20a} does not allow the agents to express certain outcomes as unacceptable due to which individual rationality is a non-issue. When agents are allowed to express certain plots or neighbors as unacceptable, individual rationality is a meaningful property. For example,  agents should not be forced to get a plot which does not fit their needs. Another scenario in which individual rationality is important is when agents already have default plots (also referred to as initial endowments) and the goal is to enable mutually beneficial exchanges. In these cases, an agent would want to get an outcome that is at least as preferred as the default outcome. 

We revisit the model of \citet{EPTZ20a} while taking into account the additional issue of individual rationality. 
We examine the challenge of designing mechanisms that simultaneously satisfy individual rationality, Pareto optimality, and strategyproofness. We present two  impossibility results that show that even under severe restrictions, the three properties are incompatible. The first impossibility result holds when agents are only allowed to express certain plots as unacceptable. The second impossibility result holds when agents are only allowed to express certain neighbors as unacceptable. We then show that the three properties are independent in our context by proving that any pair of properties can be simultaneously satisfied.

\paragraph{Related Work}

The problem of allocating indivisible items among agents is well-studied in economics and computer science. It falls under the umbrella of allocation and matching problems that take into account the preferences of the agents~\citep{Manl13a}.
The setting that we consider is based on a paper by \citet{EPTZ20a} that they refer to as `land allocation with friends'. The model we consider is slightly more general as it allows agents to express negative valuations or at least express certain plots or neighbors as unacceptable. \citet{EPTZ20a} focussed on the complexity of computing welfare maximizing allocations. They also presented a polynomial-time algorithm that is Pareto optimal and strategyproof under certain restrictions. They did not consider the issue of agents expressing certain alternatives as unacceptable. 

The aspect of preferences over neigbors makes the problem a strict generalization of the well-known house allocation problem~(see e.g., \citet{AbSo98a,Sven99a}). Since agents care about who their neighbor is, the problem also has connection with hedonic games~(see e.g.~\citet{AzSa15a}). Under particular plot topologies, the model we study also has connections with room-roommate games \citep{ABH11c,CHL+16a}.

\section{Preliminaries}

\paragraph{Model}

The problem we consider involves a 
set of agents $N = \{1, \ldots, n\}$ each of whom need to be allocated at most one of the $m$ plots in $\V = \{v_1, \ldots, v_m\}$.  
The goal is to find an allocation $\A : N \rightarrow \V$ that allocated plot $\A(i)$ to agent $i$. We allow an agent to not get any plot. 

The layout of the plots is captured by a plot graph 
$\G =\tup{\V,\E}$: which is an undirected graph where \emph{neighboring} plots $w$ and $v$ are connected by an edge $\{w,v\} \in \E$. 
Each agent $i\in N$ has a {\em valuation function} $u_i:\V\to {\mathbb Q}$: $u_i(v)$ is the value $i$ derives from receiving plot $v$. 
Agents also care about who they live next to. The relationship graph $\tup{N, F}$, where
$(i, j)\in F$ indicates that $i$ and $j$ know each other and the edge weight $\phi_{i, j}\in\Q$ is the additional utility $i$ obtains for living next to $j$. For a given statement $b$, we denote with $\mathbb{I}(b)$ the indicator function with input $b$ . The output takes value $1$ if $b$ is true and $0$ if $b$ is false.

The {utility} $U_i(\A)$ of agent $i$ under allocation $\A$ is
\begin{equation}
u_i(\A(i)) + \sum_{(i, j)\in F} \phi_{i,j}\times \mathbb{I}\left(\{\A(i),\A(j)\} \in \E\right).\label{eq:utility}
\end{equation}
We will suppose that the utility of being unmatched is zero.


\paragraph{Properties}

An allocation  is \emph{Pareto optimal (PO)} if there exists no other allocation that each agent weakly prefers and at least one agent strictly prefers. 
Individual rationality (IR) is defined as follows. 
An allocation $A$ is \emph{individually rational (IR)} if $U_i(\A)\geq 0$ for all $i\in N$. A mechanism is IR if it returns an allocation that gives utility at least zero to each agent.
Finally, an allocation mechanism is \emph{strategyproof (SP)} if no agent has an incentive to misreport her utilities functions to obtain a better outcome with higher utility. 
	Two restricted forms of IR are as follows (1) agents express certain plots to be unacceptable ($-\infty$ value) or (2) agents express certain neighbors to be unacceptable ($-\infty$ value).

\section{Impossibility of Achieving IR, PO, and SP}


We show that even when agents are only allowed to express certain plots/houses as unacceptable, there exists no IR, PO, and SP  mechanism.

\begin{theorem}
If agents are allowed to express plots as unacceptable, then there is no mechanism that is PO, IR, and SP.
		\end{theorem}
		\begin{proof}
	Consider an instance with agents $1,2,3,4,5$ and plots $v_1, v_2, v_3,v_4,v_5$, where $\E=\{\{v_1, v_2\}\}$. Agents' plot valuations are shown below, and $\phi_{1, 2}=\phi_{2,3} = \phi_{3,1}=1$ and $\phi_{1, 3}=\phi_{3,2} = \phi_{2,1}=0.2$.
		\begin{center}$I_1$=
		\renewcommand{\arraystretch}{1}
		  \begin{tabular}{  l  c  c  c c c c }
		    \toprule
		            & $v_1$ & $v_2$ & $v_3$ & $v_4$ & $v_5$\\
		    \midrule
		    agent 1 & .3     & .1    &1 &0 & 0  \\
		    agent 2 & .3     & .1    &1 &0 &0  \\
		    agent 3 & .3     & .1   &1 &0 &0  \\
		agent 4 &0.1 & 0.1 & 0.1 & 0.1& 0.1\\
		agent 5 &0.1 & 0.1 & 0.1 & 0.1& 0.1\\
		    \bottomrule
		  \end{tabular}
		\end{center}

Since the instance is symmetric, without loss of generality, let the outcome be $A=\{\{1,v_1\}, \{2,v_2\}, \{3,v_3\}, \{4,v_4\}, \{5,v_5\}\}$. All other PO outcomes are symmetric in which two of agents $1, 2, 3$ are paired up in $v_1$ and $v_2$ and third gets $v_3$. Agent 4 and 5 who are symmetric then take plots $v_4$ and $v_5$.
Note that $u_1(A)=1.3$, $u_2(A)=0.3$ and $u_3(A)=1$. Suppose agent $2$ misreports her valuation under $I_2$ as follows by reporting $v_2$ as unacceptable. Then SP requires that agent $2$ should not get utility more than $0.3$ under $I_2$.

		\begin{center}$I_2$=
		\renewcommand{\arraystretch}{1}
		  \begin{tabular}{  l  c  c  c c c c }
		    \toprule
		            & $v_1$ & $v_2$ & $v_3$ & $v_4$ & $v_5$\\
		    \midrule
		    agent 1 & .3     & .1    &1 &0 & 0  \\
		    agent 2 & .3     & --    &1 &0 &0  \\
		    agent 3 & .3     & .1   &1 &0 &0  \\
			agent 4 &0.1 & 0.1 & 0.1 & 0.1& 0.1\\
			agent 5 &0.1 & 0.1 & 0.1 & 0.1& 0.1\\
		    \bottomrule
		  \end{tabular}
		\end{center}

We note the following for instance $I_2$. 
\begin{enumerate}
	\item If agent 2 gets $v_2$, the mechanism is not IR as the agent gets an unacceptable plot.  
	\item If agent 2 gets $v_3$, the mechanism is not SP as agent $2$ can misreport under instance $I_1$.
\item If agent 2 gets $v_1$, then PO requires that agent 1 or 3 get $v_2$. Therefore, agent 2 gets utility at least $0.5$ which violates SP. 
\item From the above three cases, it is clear that either 2 is unmatched or gets $v_4$ or $v_5$. 
	\item If $1$ and $3$ take $v_1$ and $v_2$ or $v_2$ and $v_1$, then PO requires that agent $2$ gets $v_3$ which violates SP. 
	\item Suppose one of agents 1 or 3 gets $v_3$. 
	Then the other must get $v_1$ because of PO. 
	\begin{enumerate}
		\item Suppose $1$ gets $v_1$, $4$ or $5$ gets $v_2$ and 3 gets $v_3$. Then, agent $1$ gets utility $0.3$. We will show that in this case the mechanism is not PO. Agent $1$ will get utility $1.1$ if she takes $v_2$ and agent $2$ gets $v_1$ which results in a Pareto improvement hence showing the mechanism is not PO.
		\item   Suppose $3$ gets $v_1$, $4$ or $5$ gets $v_2$ and 1 gets $v_3$. Then, agent $3$ gets utility $0.3$. We will show that in this case the mechanism is not PO. Agent $3$ will get utility $0.3$ if she takes $v_2$ and agent $2$ gets $v_1$ which results in a Pareto improvement hence showing the mechanism is not PO.
	\end{enumerate}
\end{enumerate}

We have exhausted all the cases and found that in each case, either IR, SP, or PO is violated. 
\end{proof}

Next, we show that if agents are allowed to express certain neighbors as unacceptable, then there is no mechanism that is PO, IR, and SP. Our proof
relies on connecting the problem with marriage markets in particular exploiting a known impossibility result by \citet{AlBa94a}.


\begin{theorem}
If agents are allowed to express certain agents as unacceptable,
then there is no mechanism that is PO, IR, and SP.
	\end{theorem}
	\begin{proof}
Consider a generic instance with agents $m_1,m_2,w_1,w_2$ and plots $v_1, \ldots, v_8$, where $\E=\{\{v_1, v_2\}, \{v_3, v_4\},\{v_5, v_6\}, \{v_7, v_8\}\}$.  
We assume that agents get zero value for each of the plots. 
For reasons that will become clear immediately, we refer to $m_1$ and $m_2$ as men and $w_1$ and $w_2$ as women.
We suppose that men have a sufficiently large negative value for men and women have a sufficiently large negative value for women. Agents have strictly positive value for the members of the opposite gender unless they express them as unacceptable. 




For the input above, an agent only cares about who the agent is paired with or whether it is alone. We partition the set of possible allocations for the problem $I$ into classes where each class of outcomes has the same pairing between agents. Each agent is indifferent between allocation in the same class.
Among the classes of allocations, we call a class IR if the allocations in the class are IR.  

Next, we present a reduction from the problem of land allocation with neighbors to the marriage market problem as follows. 
For a generic instance $I$ of the form above, we can construct a corresponding instance $I'$ of a marriage market involving two men and two women in which men have preferences over women and women have preferences over men. 
In particular, for instance $I'$ the preferences of men over women and of women over men is the same as in $I$. In problem instance $I'$ there are no plots. For a marriage market, an outcome is a matching between men and women.

Next, we prove the following claim. 
\begin{claim}
	There is a one-to-one correspondence between IR classes of allocations for $I$ and matchings for $I'$. 
	\end{claim}
	\begin{proof}
For a given IR equivalence class of allocations for instance $I$, there is a corresponding matching for instance $I'$ in which a man and women are matched for $I'$ if they were neighbors for the allocation under $I$. Coversely, for a matching for $I'$, there is a corresponding class of allocation for $I$ in which matched man-woman pairs are neighbors. 
We have established a one-to-one correspondence between IR classes of allocations for $I$ and matchings for $I'$.
\end{proof}

We note that in the one-to-one correspondence between IR classes of allocations for $I$ and matchings for $I'$, agents have the same preferences over the corresponding outcomes. 
\citet{AlBa94a} in the proof of their Proposition 1 showed that for marriage problem with two men and two women, there exists no IR, PO, and strategyproof mechanism. The same statement also follows for our setting because we have designed a reduction from $I$ to $I'$ for which our claim holds: there is a one-to-one correspondence between IR classes of allocations for $I$ and matchings for $I'$.
If we wish to keep the number of agents the same as the number of plots, we can create four dummy agents who do not derive or provide value by being neighbors.
\end{proof}

We conclude with three remarks.

\begin{remark}[Cardinal versus Ordinal Preferences]
We note that we have framed the setting in a manner that is consistent with that of \citet{EPTZ20a}. However, the model that we consider as well as the results do need to assume additive utilities. They also hold if agents have ordinal preferences over outcomes. 
\end{remark}

\begin{remark}[Independence of the axioms]
Each pair of the properties from IR, PO, and SP are compatible. An IR and PO outcome exists by starting from an IR outcome and repeatedly getting Pareto improvements until no further Pareto improvement can be achieved. SP and IR are compatible because there is an option to not allocate any plot. SP and PO are also compatible via the generalized serial dictatorship mechanism studied by \citet{ABH11c}.
\end{remark}

\begin{remark}[Computation]
We have not discussed computational issues up till this point. We note that computing a Pareto optimal outcome is NP-hard even if the plots are in pairs and agents have additively separable 1-0 utilities over plots and over neighbors. The argument is as follows. \citet{CHL+16a} proved that maximizing social welfare for room-roommate games is NP-hard even if agents have additively separable 1-0 utilities over rooms and over roommates. If one examines the reduction, it also proves that even checking whether each agent gets a most preferred outcome is NP-complete. It follows then from Lemma~1 of \citet{ABH11c} that computing a PO outcome is NP-hard.
	\end{remark}

	\bibliographystyle{plainnat}
	




	\end{document}